\documentclass{ifacconf}

\usepackage{graphicx}
\usepackage{graphicx}      
\usepackage{natbib}        
\usepackage{amssymb}

\usepackage{amsmath}
\usepackage{tikz}
\usetikzlibrary{backgrounds,arrows,automata,fit,arrows,decorations.pathreplacing,calc,positioning,decorations.text,fadings,shapes.arrows,shadows}
\usepackage{comment}
\usepackage[short]{optidef}

\usepackage{lineno}

\usepackage{blindtext}

\usepackage{todonotes}

\usepackage{macro}

\usepackage{algpseudocode}

\usepackage{booktabs}

\usepackage{mhchem}

\usepackage{siunitx}

\theoremstyle{definition} 
\newtheorem{definition}{Definition}
\theoremstyle{plain} 

\theoremstyle{remark} 
\newtheorem{remark}{Remark}
\theoremstyle{plain} 

\begin{document}
\begin{frontmatter}
\title{Occupation measure methods for modelling and analysis of biological hybrid systems} 

\author[First,Second]{Alexandre Rocca}
\author[First]{Marcelo Forets}
\author[First]{Victor Magron}
\author[Second]{Eric Fanchon} 
\author[First]{Thao Dang} 

\address[First]{Univ. Grenoble Alpes, CNRS, Grenoble INP, VERIMAG, 
   700, avenue centrale,38000, Grenoble, France}
\address[Second]{Univ. Grenoble Alpes, Grenoble 1/CNRS, TIMC-IMAG, UMR 5525,
				38041, Grenoble,France \thanksref{emails}}

\thanks[emails]{For all authors emails follow forename.surname@univ-grenoble-alpes.fr}

\begin{abstract}
	Mechanistic models in biology often involve numerous parameters about which we do not have direct experimental information. 
The traditional approach is to fit these parameters using extensive numerical simulations (e.g. by the Monte-Carlo method), and eventually revising the model if the predictions do not correspond to the actual measurements. 
In this work we propose a methodology for hybrid system model revision, when new types of functions are needed to capture time varying parameters.
To this end, we formulate a hybrid optimal control problem with intermediate points as successive infinite-dimensional linear programs (LP) on occupation measures. Then, these infinite-dimensional LPs are solved using a hierarchy of semidefinite relaxations.
The whole procedure is exposed on a recent model for haemoglobin production in erythrocytes.

\end{abstract}

\begin{keyword}
biological modelling, hybrid dynamical system, optimal control problem, semidefinite optimization, occupation measures.
\end{keyword}

\end{frontmatter}

 \section{Introduction and Context}
 \label{sec:intro_main}

Mechanistic models in biology generally involve many parameters. The value of a given parameter can be either measured directly in a dedicated experiment (e.g. measurement of a kinetic parameter of a biochemical reaction in enzymology), or inferred from data which provide relationships between parameters and other biological entities.

In this paper, we work with biological mechanisms modelled by ordinary differential equations (ODEs) and hybrid dynamical systems. 
The motivation for using such models is to give quantitative predictions, when sufficient data is available to validate the model.
And whenever new data and knowledge of various types become available, they can be incorporated within a formal framework.

A basic issue in biological systems modelling is the determination of numerical values for the parameters, or more generally a subset of the parameter space, under which the model agrees to some extent with the available data. 
We focus on multiple-step experiments, in which a biological system is perturbed or measured during its evolution. 

In the biological systems modelling literature, it is common to synthesise parameters using a Monte-Carlo sampling of the parameter space, which is validated then by numerous simulations. 
An important effort to formalize and validate the parameter synthesis of biological models has been made in works such as \cite{breachHybrid,breachMobilia,sapo,benevs2016parallel}, or \cite{paraInvalidation}. Other such as \cite{cardelli2017syntax} or \cite{bartocci2013temporal} design ODE models satisfying sets of temporal constraints. 
When model simulation does not reproduce satisfactorily available experimental data, to a degree which depends on data quality, for any admissible parameter value, the model has to be revised. 
One way of revising the model is to represent parameters using different types of functions of time, reflecting underlying biological mechanisms. 
We introduce a systematic way, based on formal methods, to study mechanistic biological models in their experimental context and revise parameters to produce conservative results with respect to experimental data. 
In this work, we consider a problem of model revision, defined as finding time varying laws of parameter evolution that minimizes the error in matching experimental measurements. Concretely, it is the following optimization problem:
\begin{mini}
	{(\bx,\bu)}{\sum_{j=1}^{\nbexp}\text{dist}(\bm(\bx(T_j)), \bz_j)}
	{\label{eq:ocpintro}}{}
\end{mini}
where $\bx$ is a vector of biological variables, such as concentrations, whose evolution is modelled by trajectories of a biological dynamical system.
Time varying parameters are represented by the input variables $\bu$ (modelling biological parameters) such that  $\forall t \in [0,T] \,, \bu(t) \in \InputSet$. 
$\X_0$ is the set of initial conditions and the set of pairs $\{(T_j, \bz_j)\}_j$ is the set of data points, for $1 \leq j \leq \nbexp$, in the time frame $[0, T]$. 
An experimental measurement is a function of the variables $\bx$ and is modelled via the function $\bm(\bx)$. 

The framework of our approach is a mathematical formalization of experimental protocols as hybrid dynamical systems, describing biological systems of interest and experiments which are performed on them. 
However, the algorithm we provide can be applied to any biological hybrid system with similar model revision problems.

In this paper we address the parameter synthesis and model revision problems \eqref{eq:ocpintro} by formulating  a particular instance of the optimal control problem with intermediate costs, when the objective function depends on the systems trajectory and control inputs at a given set of time points. 
This problem is then approximated by multiple hybrid optimal control problems (HOCP) with one final cost.  
To this end, we apply a recently developed method of \cite{hybridOCP} from the field of certified convex optimization to globally solve these HOCP. 
However, The method described in \cite{hybridOCP} produces piecewise optimal control functions which either may not correspond to biological knowledge of parameter variations or may be difficult to yield coherent and meaningful biological interpretations. 
Consequently, in order to respect realistic constraints on parameters, we use smooth approximations of the generated control input, in order to revise the given model while maintaining good data fitting accuracy. 

The method is demonstrated on a hybrid system modelling haemoglobin production presented in Section \ref{ssec:res}.

 	\subsection{Related work}
     \label{sec:intro_relat}
 		The hybrid formalism has previously been used as an abstraction method to simplify complex mechanisms which are hard to analyse as seen in \cite{cellcyclehybrid,iron_homeostasis}, or to represent ``jump'' evolution such as activation processes in genes regulatory networks for example using the stochastic formalism as in \cite{review_stochasticHS}. 

Optimal control theory and variation theory have been applied to biological systems in several works. 
Most of them address the classical problem of finding a correct input such that the system reaches a desired state. For example, one can control drug input such that a patient attains a healthy state, see \cite{drug_control} or \cite{caraguel2016towards}. Another example is the control of some input in population studies as detailed in \cite{population_studies}. A detailed review on the use of optimal control in systems biology can be found in \cite{book_OptCtrlSysBio}. The problem of parameter estimation in presence of multiple data, also called data assimilation, is stated in \cite[Chapter 26]{book_OptCtrlSysBio}. 

The optimal control problem for specific classes of hybrid systems has been investigated in several domains, such as mechanical systems in \cite{gradientHOCP} and switched-mode systems in \cite{wardi2015switched,xu2004optimal,bengea2005optimal}. 
More generally, \cite{PontriandDynamic} relies on Dynamic Programming and an extension of Pontryagin's Maximum Principle. 
However, these approaches need a-priori knowledge either on the sequence of discrete transitions, or on the number of visited subsystems. 
To perform optimal control on hybrid systems, we build our work on the techniques from \cite{hybridOCP}, which proposes a method to obtain a global solution for hybrid systems with state-dependent transitions, without any a-priori knowledge on the execution and the sequence of transitions. 
We refer to \cite[Section 1.1]{hybridOCP} and references therein for more details on optimal control of hybrid systems.

Semidefinite programming (SDP) eases the resolution of hard optimization problems and yields conservative results ensured by positivity certificates. In \cite{polyopti}, hierarchies of semidefinite relaxations were introduced for static polynomial optimization. 
The definition of an infinite-dimensional linear program (LP) over occupation measures, for optimal control problems, was first introduced in \cite{vinter}. 
From this infinite-dimensional LP, \cite{ocp} defines hierarchies of Linear Matrix Inequalities (LMI) relaxations, to synthesise a sequence of polynomial controls converging to the solutions of the optimal control problem. 
In \cite{ocppwa} the authors propose an extension to piecewise affine systems. 
Our underlying idea of constructing a suboptimal control with an iterative algorithm is similar to \cite[Section 4]{ocppwa}. 
However, we use this scheme to find input functions allowing to reproduce data not only at a final time point but also at intermediate time points. 
We make use of the recent method proposed in \cite{hybridOCP}, which relies on occupation measures and a sequence \emph{semidefinite relaxations} to produce a sequence of polynomial controls converging to the optimal solution of a HOCP.
%
%
 %
There exist other methods which use occupation measures and LMI relaxations to produce both admissible controls and converging outer-approximations of the backward reachable set (BRS) \cite{brs,hybridBRS}, or the region of attraction (ROA) \cite{roac}.
Finally, we note that finding a sequence of converging outer-approximations for all valid parameters sets, such as in \cite{param_estim, sapo}, is another crucial issue in the context of systems biology. 
When dealing with hybrid systems, \cite{hybridBRS} can be applied to solve this problem.

 	\subsection{Main contributions}
      \label{sec:intro_contrib}
 		Given a hybrid system modelling a multiple-step biological protocol and a set of experimental measurements, we propose a numerical algorithm to solve the model revision problem. 
This algorithm generates an admissible input function such that the revised model reproduces the experimental measurements, and can be interpreted.
We release a software package\footnote{https://github.com/biosdp/biosdp}, in  MATLAB, implementing this algorithm. We evaluate this algorithm and its implementation, on the haemoglobin production model taken from \cite{hemeModel}.

The rest of the paper is organized as follows: in Section \ref{sec:prelim} we give the necessary notations on dynamical hybrid systems and the optimal control problem. Section \ref{S:opti} presents our main contribution to the resolution of the hybrid optimal control problem with intermediate points. Finally, in Section \ref{S:result} we study and discuss the model revision of the haemoglobin production model taken from \cite{hemeModel}.
        
 \section{Preliminaries}
 \label{sec:prelim}
     We first give the notations and basic notions on controlled hybrid systems, as well as the definition of the HOCP. 


Given $\bx \in \R^{n} $, let $x_i$ denote its i-th component. 
In general, letters in bold font denote multidimensional elements, and normal font unidimensional ones. 
Let $\mathbb{B} := \{ \texttt{true}, \texttt{false}\}$ be the set of Booleans.
Let $\R[\bx]$ denote the ring of real polynomials in $\bx\in \R^{n} $, and let $\R_d[\bx]$ be the subspace of polynomials whose degree is at most $d$. 
Let $\Tint$ be the time interval $[0,T]$, where $T$ is the final time (possibly $\infty$).

Consider the $n$-dimensional ODE with inputs:
\begin{equation}
    \label{eq:ode}
	\dot{\bx}(t) = \mathbf{f}(t,\bx(t),\bu(t)),
\end{equation}
with $\mathbf{f}: \Tint \times \R^n \times \R^{\dimInput} \to \R^n$ a vector field which is Lipschitz continuous in $\bx$ and piecewise continuous in $\bu$.
Let $\X$ and $\InputSet$ be compact subsets of $\R^n$ and $\R^{\dimInput}$ respectively. 
Here, $\bu:~ \Tint \to \InputSet$ is a feasible input function 
which represents time varying parameters, or external commands. 
The tuple
	\begin{equation}\label{eq:dy_sys}
			\mathcal{F} := (\Tint, \X, \InputSet, \mathbf{f})
	\end{equation}
defines a continuous dynamical system.

We recall the definition of \emph{controlled hybrid systems} a dynamic hybrid systems formalism defined in \cite{hybridOCP}.
\begin{definition}[Controlled hybrid system]\label{def:HybridAutomaton}
A controlled hybrid system (CHS) is defined by the tuple: $\Hybrid = (\I, \Trans, \UDom, \InputSet, \F, \Guard, \Resetmap)$ where:
\begin{itemize}
    \item $\I \subset \Nset$ is the set of mode indices, and $n_{modes}$ the number of modes.
    \item $\Trans \subseteq \I \times \I$ is the set of transitions $\tr = ({\sti},{\stj})$ between two modes: ${\sti}$ is the source mode, and ${\stj}$ the destination mode.
    \item $\X:= \coprod_{i\in\I} \X_{\sti}$ is the disjoint union of domains of $\Hybrid$ and $\X_{\sti}$ the domain of the mode ${\sti}$. We note that  $\X_{\sti}$ is a compact subset of $\R^{n_i}$ with $n_i$ the dimension of the mode $i$. The disjoint union $\coprod$ can simply be considered as a labelling operation on the set of domains by $\I$, that is the set of mode indices.
 	\item $\InputSet$ is the set of input values of $\Hybrid$.
    \item $\F := \{\F_{\sti}\}_{{\sti}\in\I}$ is the set of continuous dynamical sub-systems associated to each mode. The dynamical system associated to mode ${\sti}$ is: 
    $$\F_{\sti} := (\Tint,\X_{\sti},\InputSet,\mathbf{f}_{\sti})\,,$$
    with $\mathbf{f}_{\sti}: \Tint \times \X_{\sti} \times \InputSet \to \R^{n_i}$ a vector field polynomial in $\bx$ and affine in $\bu$.
    \item $\Guard := \coprod_{\tr\in\Trans} \Guard_{\tr} $ is the disjoint union of guards $\Guard_{\tr}\subseteq\X_{\sti}$ associated to each transition $\tr = (\sti,\stj) \in \Trans$. The guard $\Guard_{(\sti,\stj)}$ defines the switch condition from $\sti$ to $\stj$: for $\bx\in \X_{\sti}$, if $\bx\in\Guard_{(\sti,\stj)}$ then the system at $\bx$ can make the transition from mode $\sti$ to mode $\stj$.
    \item $\Resetmap := \{\Resetmap_{\tr}\}_{\tr\in\Trans}$ is the set of reset maps, each reset map $\Resetmap_{\tr} : \Guard_{\tr} \to \X_{\stj}$ being associated to a transition $\tr := (\sti,\stj) \in \Trans$ and it defines how the continuous variables may change after the discrete transition from mode $\sti$ to mode $\stj$.
\end{itemize}
\end{definition}

Additionally, the CHS defined in \cite{hybridOCP} must respect a few assumptions. 
All the guards $\Guard_{(\sti,\cdot)}$ are disjoint, and $\Guard_{(\sti,\stj)} \subseteq \partial \X_\sti$, with $\partial \X_\sti$ designating the border of  $\X_\sti$, for each pair of modes $\sti$ and $\stj$. 
The initial set is restricted to a single point $\bx_0$, with an associated mode $i_0$. The vector fields $\mathbf{f}_{\sti}$ are affine in $\bu$ and have a nonzero normal component on the boundary of $\X_\sti$. 
These assumptions ensure that any CHS is deterministic.
Noting $\lambda(\bx(t))$ the function which associates to an instantaneous state $\bx(t)$ its corresponding mode, these assumptions ensure that the mode corresponding to $\bx(t)$ is unique.

Given a CHS $\Hybrid$, the hybrid optimal control problem is defined as follows.
Let $\X_0$, and  $\X_T$, be the initial set, and target set defined by:
 \begin{equation}
 	\X_0 := \coprod_{i\in\I} \X_{0,i}\,, \quad \X_T := \coprod_{i\in\I} \X_{T,i}\,,
 \end{equation}
where  ${\X_0}_i$ and ${\X_T}_i$ are a compact subsets of $\X_i$ for each mode $i\in\I$. 
Let $i_0$ and $i_T$ be the initial mode and the final mode at time $T$, respectively. 
Then, given $(i_0,\bx(0)) = (i_0,\bx_0) \in \X_0$ and $\bu:\Tint\to\InputSet$ an input function, we say that for $T>0$, $(\bx(t),\bu(t)) \in \admiPair$ is an admissible pair on $\Tint$ and $\admiPair$ is the set of admissible pairs, if $(i,\bx(t))\in \X$ is a trajectory of CHS as defined by \cite[Algorithm 1]{hybridOCP} and $(i_T,\bx(T)) \in \X_T$. 
Finally, the hybrid optimal control problem for a CHS $\Hybrid$, (HOCP), is defined by:
\begin{equation}\label{eq:hocp}
\begin{aligned}
J_{hocp}^* :=~~ \inf_{(\bx,\bu)\in\admiPair} & \int_0^T h_{\lambda(\bx(t))} \left(t, \bx_{\lambda(\bx(t))}(t), \bu(t)\right)dt 
\\ 
				& \qquad + H_{\lambda(\bx(T))}\left(\bx_{\lambda(\bx(T))}(T)\right)\,.
\end{aligned}
\end{equation}
where $\{h_i : [0, T]\times \R^{n_i}\times \R^m \to \R\}_{i\in \joli{I}}$ and $\{H_i : \R^{n_i} \to \R\}_{i\in \joli{I}}$ are set of measurable functions. 
 \section{Optimal control for model revision}
 \label{S:opti}
 	In our biological context hybrid dynamical systems can model multiple steps experiments. 
Indeed, experimental protocols associated to these experiment can be considered as a set of concurrent processes where each process is modelled as a hybrid system. 
Thus, the hybrid system of the protocol can be described as the parallel composition of the hybrid system describing each process. 
It is then crucial to proceed to model revision while taking into account the biological system in the evolving environment of the complete protocol. 
For this purpose, we propose in this section a method to produce time varying parameters reproducing multiple data in the context of a multiple-phase protocol modelled by a hybrid system. 
However, we also argue that this method can be use for more general biological systems modelled as CHS.
We solve the model revision problem of a dynamical hybrid system modelling a biological system together with a set of experiments. 
Therefore, we search for parameters as time varying functions fitting a set of data points, defined in the introduction as in \eqref{eq:ocpintro}.
In this aim, Section \ref{ssec:chs} first formulates \eqref{eq:ocpintro} as a particular instance of the optimal control problem on hybrid systems with intermediate costs. 
Then, we propose a first approximation as a set of instances of the HOCP \eqref{eq:hocp} defined previously. 
The solution of each sub problem is obtained using the previous results from \cite[Section 4]{hybridOCP}.
Finally, in Section \ref{ssec:implementation} we explain the complete algorithm addressing our initial problem.



 	\subsection{Problem statement}\label{ssec:chs}
We provide a method to find time varying parameters of biological system, modelled as input functions $\bu(t)$, in order to fit the hybrid systems model to a set of experimental data.
Thus, we write our problem as an optimal control problem where desired input functions are the optimal controls which minimize the distance of the results produced by the model and these experimental data.
Experimental measurements, represented by a function $\bm(\bx)$, are performed at given specific times $T_j$, $1 \leq j \leq \nbexp$. 
Let $\bz_{j}$ be the observed value of the experimental measurement at time $T_j$, then $\nbexp$ is the number of experimental data points. 
Let $\X_{T_{j,i}}$ be compact subsets of $\X_i$, and  $\X_{T_j} := \coprod_{i\in\I} \X_{T_j,i}$. As in \eqref{eq:hocp}, let $(i_0,\bx(0)) \in \X_0$, and suppose that we are given a set of time values $\{T_{j}\}$, with $1\leq j \leq \nbexp$, and $T_{\nbexp} = T$. Given an input function $\bu:\Tint\to\InputSet$, we say that $(\bx,\bu)\in \admiPair_{int}$ is an admissible pair for a problem with intermediate points and $\admiPair_{int}$ the associated set of admissible pairs, if $(i(t),\bx(t))\in\X$ is a CHS trajectory, and $(i_{T_j},\bx(T_j)) \in \X_{T_j}$  for all $j$. 

Let $H(\bx(T_j))$ be a cost at time $T_j$, and $h(t,\bx(t),\bu(t))$ a running cost for the whole $[0,T]$ interval. The optimal control problem with intermediate points for the CHS $\Hybrid$ is then:
\begin{equation}\label{eq:Int_HOCP}
\begin{aligned}
J_{hocp}^* :=~~ \inf_{(\bx,\bu)\in\admiPair_{int}} & \int_0^T h(t,\bx(t),\bu(t)) dt \\
	& + \sum_{0\leq j \leq \nbexp}H\left(\bx(T_j)\right)
\end{aligned}
\end{equation}
In our biological context, $ H\left(\bx(T_j)\right)  = || \bm(\bx(T_j)) - \bz_{j}||_2^2.$

To our best knowledge there is no method to efficiently address directly problem \eqref{eq:Int_HOCP}, and obtain converging sequence of solution in a similar manner to the simpler problem \eqref{eq:hocp} address in \cite{hybridOCP}. 
Thus, we search for an admissible solution using a greedy algorithm. 
Consequently, this solution is a good trade-off between computation cost, optimality and flexibility as we can handle a large class of models. 
Moreover, this method does not constraint the form of the sought times parameters: this reduces the assumptions during the modelling and eases the final biological interpretations, unlike methods based on simulations.
Given $1 \leq j \leq \nbexp$, let
\begin{equation*}
J_j(t,\bx(t),\bu(t))  := \int_{T_{j-1}}^{T_j} h(t,\bx(t),\bu(t)) + H\left(\bx(T_j)\right),
\end{equation*}
with $T_{0} = 0$, and $T_{\nbexp} = T$, such that 
\begin{equation*}
J(t,\bx(t),\bu(t)) = \sum_{1\leq j \leq \nbexp}  J_j(t,\bx(t),\bu(t)).
\end{equation*}
Noting $(i^{(j)}(t),\bx^{(j)}(t))$ a trajectory of a CHS $\Hybrid$ on the interval $\Tint_j := [T_{j-1},T_j]$, and similarly $\tilde{\bu}^{(j)}(t)$ the control on $\Tint_j$, we consider the following problem as particular instance of \eqref{eq:hocp}:
 \begin{equation}\label{eq:pHOCP}
 \begin{aligned}
 J_j^* :=~~& \underset{(\bx^{(j)},\tilde{\bu}^{(j)})}{\inf}   J_j(t,\bx^{(j)}(t),\tilde{\bu}^{(j)}(t)) \\
 \text{  s.t.} & \\
 & \bx^{(j)} \textnormal{ cont. part of a trajectory } (i^{(j)},\bx^{(j)})  \textnormal{ on } \Tint_j,\\
 & \tilde{\bu}^{(j)}(t) \in \InputSet\,, \quad \forall t\in \Tint_j\,,\\
 & (i^{(j)}(t),\bx^{(j)}(t)) \in \X \,,\quad \forall t\in \Tint_j\,,\\
 & \textnormal{ if } j=1\,, \\
 & \quad \, (i^{(1)}(0),\bx^{(1)}(0)) \in \X_0\,,\\ 
 & \textnormal{ if  }j \geq 2.\\
 & \quad \, (i^{(j)}(T_{j-1}),\bx^{(j)}(T_{j-1})) = (i^{(j-1)}(T_{j-1}),\bx^{(j-1)}(T_{j-1}))\,, \\ 
 & (i^{(j)}(T_j),\bx^{(j)}(T_j)) \in \X_{T_j}.
 \end{aligned}
\end{equation}

We note that if a transition $i \to i'$ occurs at the time $T_j$ of the interval $[T_{j-1},T_j]$, we retain only the left part in the mode $i$ for the next optimization on the interval $[T_{j},T_{j+1}]$.
Let $\tilde{\bu}(t)$ and $(i(t),\bx(t))$ be respectively the control and the trajectory, for $t\in [0,T]$. They are respectively defined by the concatenation of all the controls $\tilde{\bu}^{(j)}(t)$ and the trajectories $(i^{(j)}(t),\bx^{(j)}(t))$ on the sub-intervals $[T_{j-1},T_j]$. By construction, $(\bx(t),\tilde{\bu}(t))$ is an admissible pair for \eqref{eq:Int_HOCP}, as $(i_{T_j},\bx(T_j)) = (i^{(j)}_{T_j},\bx^{(j)}(T_j)) \in X_{T_j}$.
\begin{remark}\label{rmk:opti}
We emphasize that $(\bx(t),\tilde{\bu}(t))$ is not necessary an optimal solution for \eqref{eq:Int_HOCP}. Moreover, as  the optimization problem \eqref{eq:pHOCP} is obtained through a greedy scheme, we have no guarantee that its optimal cost $ J_j^*$ is inferior to a given $\varepsilon$. However, as we search to equally fit all the data points searching iteratively for the control is satisfiable solution.
\end{remark}
      \subsection{Implementation}\label{ssec:implementation}

Let $(T_j,\bz_{j})$,  $0 \leq j \leq \nbexp$ be pairs of experimental data points and their measurement time, and we also note $i_0$, and $\bx_0$ the initial mode and initial conditions of the studied CHS $\Hybrid$ respectively. 
Let $r$ be a given starting relaxation degree. Algorithm \ref{alg:pw_ocp} finds an admissible solution to \eqref{eq:Int_HOCP}, by solving a sequence of the HOCP \eqref{eq:pHOCP} for each experimental data point $(T_j,\bz_{j})$. 
For each $j$, the degree of the polynomial control $\tilde{\bu}^{(j)}(\bx(t),t)$ is determined as the smallest degree such that  $|| \bm(\bx(T_j)) - \bz_{j}||_2^2 \leq \varepsilon$. 
Indeed, in the context of biological system modelling we desire to obtain a control of degree as small as possible to avoid overfitting.
We note that as explained in Remark \ref{rmk:opti}, we cannot ensure the converge to a solution of accuracy $\varepsilon$. 
Thus, $\varepsilon$ is only a stopping criteria.
Then, for each iteration over $j$, Algorithm \ref{alg:pw_ocp} is decomposed in three steps. 

The first step is the procedure \texttt{HOCP}, associated to an instance of the HOCP \eqref{eq:hocp} for $j$-th pairs $(T_j,\bz_{j})$. 
Given a relaxation order $d_r \geq r$, we solve the relaxed primal defined in \cite[Section 5.1]{hybridOCP}. 
From \cite{hybridOCP} method we first obtain $M_{d_{r}}(\by_{\mu_i})$, the sequence moment matrices of degree $d_r$, associated to the occupation measure $\mu_i$ of each mode $i\in I$. We also obtain $\underline{J}_j^{(d_r)}$ an under approximation of the optimum of \eqref{eq:pHOCP}. The second step is the procedure \texttt{Synth}, which returns the admissible control $\tilde{\bu}^{(j)}(t,\bx)$ of degree $d_{u} \leq d_{r}$ using a truncated moment matrix $M_{d_{u}}(\by_{\mu_i})$  of $M_{d_{r}}(\by_{\mu_i})$ at the reduced degree $d_{u}$. The third and last step is the procedure \texttt{Simu}. 
It performs the validation that the synthesised control $\tilde{\bu}^{(j)}$ yields $||\bm(\bx(T_j)) - \bz_{j}||_2^2 \leq \varepsilon$.
This step is done by approximating the trajectory of the controlled hybrid system using a solver of ODE with discrete events to produce numerical simulations. If in iteration $j$, $||\bm(\bx(T_j)) - \bz_{j}||_2^2 \leq \varepsilon$, then $\bx^{(j)}(T_j)$ and the corresponding mode $i_{f}$ reached at $t = T_j$ by the numerical simulations are the initial conditions for the next iteration $j+1$. Otherwise, the \texttt{Ctrl\_Synth} and \texttt{Simulate} procedures are repeated while increasing the degree of the synthesised polynomial control until $d_{u} = d_{r}$. 
In case the condition $||\bm(\bx(T_j)) - \bz_{j}||_2^2 \leq \varepsilon$ is still not satisfied, the relaxation order $d_r$ is increased, and the three steps are repeated. 

If $\varepsilon \leq \underline{J}_j^{(d_r)}$ then we are sure that we the given initial condition at step $j$, there is no control such that $||\bm(\bx(T_j)) - \bz_{j}||_2^2 \leq \varepsilon$. Consequently, we keep our previous result $\tilde{\bu}^{(j)}$ and the corresponding mode $i_{f}$ reached at $t = T_j$ by the numerical simulations are the initial conditions for the next iteration $j+1$.
\begin{algorithm}[h!]
  \caption{}
  \label{alg:pw_ocp}
  \begin{algorithmic}[1]
    \Procedure{Algorithm $1$}{$\Hybrid,\{(T_j,\bz_{j})\}_j,i_0,\bx_0,\varepsilon,r$}
    \State $T_{init} = 0$
    \ForAll{experimental data $(T_j,\bz_{j})$}
        \State $d_u = \mathbf{0}, d_{r} = r, \texttt{err} = +\infty$
    	\While{$\texttt{err} \geq \varepsilon \wedge \underline{J}_j^{(d_r)}\leq \varepsilon$}
          	\State $\underline{J}_j^{(d_r)},\mathbf{M}_{d_{r}}(\by_{\mu})$ = \texttt{HOCP}($\Hybrid,i_0,$...
          	\State \quad\quad\quad\quad\quad\quad\quad\quad\quad...$\bx_0,T_{init},T_j,\bz_{j},d_{r}$)
          	\While{$\texttt{err} \geq \varepsilon$ \textbf{and} $d_u \leq d_{r}$}
              \State $\tilde{\bu}^{(j)}(\bx(t),t) = \texttt{Synth}(\mathbf{M}_{d_{r}}(\by_{\mu}),d_u$) 
              \State $(i_{f},\bx^{(j)}(t)) = \texttt{Simu}(\Hybrid,\tilde{\bu}^{(j)}(\bx(t),t),$...
              \State \quad\quad\quad\quad\quad\quad\quad\quad\quad ... $i_0,\bx_0,T_{init},T_j)$
              \State $\texttt{err} = H(x^{(j)}(T_j),\bz_{j})$
              \State $\texttt{increase } d_u$
          	\EndWhile
        	\State $\texttt{increase } d_{r}$
        \EndWhile
        \State $i_0$ = $i_{f}$
        \State $\bx_0 = \bx^{(j)}(T_j)$
        \State $T_{init} = T_j$    
    \EndFor
    \EndProcedure
  \end{algorithmic}
\end{algorithm}

 \section{Results and Discussion}
 \label{S:result}

In this section, using the method developed in Section \ref{S:opti}, we revise the model of haemoglobin production by finding a better fit for the time varying parameter noted $k_3$ in \cite{hemeModel}, with respect to the same error function. 
To be consistent with our notations from Sections \ref{sec:prelim} and \ref{S:opti}, we will note $u(t)$ the input modelling the time varying parameter $k_3$.

\subsection{Application to haemoglobin production and results}
\label{ssec:res}

The paper \cite{hemeModel} addressed, on an experimental protocol model, the problem of refining the parameters space, and to fit multiple data sets of experimental results. 
In \cite{hemeModel}, the experimental protocol was not explicitly formalized as a hybrid system. 
However, we use the parameters value from this previous work as starting instantiation to provide less restrictive constraints on the time varying parameter $k_3$. 

The ODEs \eqref{eq:fctrl} model the evolution of the haemoglobin production in the differentiating erythrocyte cells situated in the bone marrow. The variables $x_1$ to $x_4$ represent respectively the internal iron in the cell $\ce{Fe}$, the heme $\ce{H}$ , the globin $\ce{G}$, and the haemoglobin  $\ce{Hb}$. 
The hybrid system $\Hybrid$ models an experimental protocol designed to measure the integration of iron inside heme ($\ce{H}$) at multiple milestones of the cell differentiation. For example the data point at time $t=7$ hours in Table \ref{tab:datas}, is obtained through the following procedure: we first start with a control batch of cells, then at time $t=4$ hours after the start of the experiment, the culture medium is perturbed with an injection of measurable radioactive iron $\ce{^{\scriptscriptstyle 59}Fe}$ for a subset of the cells. 
This perturbation implies new ODEs \eqref{eq:frad} modelling the evolution of two inter-dependant models: the model of non-radioactive haemoglobin production and the model of haemoglobin production with radioactive species. 
Three hours after the perturbation with radioactive iron, the total radioactive heme is measured, meaning the heme free in the cell and the one in the radioactive haemoglobin. This measurement is given by the formulas $\ce{^{\scriptscriptstyle 59}H} + 4\ce{^{\scriptscriptstyle 59}Hb}$. The data in Table \ref{tab:datas} is then the observed radioactivity divided by three hours. 
Finally, these measurements provide results on the variation during the cell differentiation of the integration of iron in heme, which is associated to the parameter $k_3$.

The controlled hybrid system $\Hybrid$ associated to the experiment studied in \cite{hemeModel} and the haemoglobin production model are given, in a shortened version, in Tables \ref{tab:hybrid} and \ref{tab:trans}. 
The ODEs \eqref{eq:fctrl} and \eqref{eq:frad}, as well as the numerical values of the parameters, are given in Appendix. 
In the implementation, we also introduce a variable $x_c$ modelling time, whose derivative is equal to $1$. 

\begin{table}
\begin{center}
	\caption{Dimensions (with $x_c$), vector fields, domains, and input sets for the controlled hybrid system $\Hybrid$ of the haemoglobin production model.}\label{tab:hybrid}
	\begin{tabular}{lclccc} \toprule
		{Mode}  & $n_i$ & $\mathbf{f}_i(t,\bx,\bu)$  &  $\X_i$                & $\InputSet_i$        \\ \midrule
		{$i=1$} & 5  & $\mathbf{f}_{ctrl}(t,\bx,u)$  & $[0,4]\times[0,1]^4$   &  $[0,1]$           \\ \midrule
		{$i=2$} & 9  & $\mathbf{f}_{rad}(t,\bx,u)$   & $[4,7]\times[0,1]^8$   &  $[0,1]$            \\ \midrule   
		{$i=3$} & 5  & $\mathbf{f}_{ctrl}(t,\bx,u)$  & $[7,8]\times[0,1]^4$   &  $[0,1]$             \\ \midrule
		{$i=\vdots$}  & $\vdots$ 		 & $\vdots$  & $\vdots$    &  $\vdots$                       \\ \midrule		
		{$i=13$} & 5 & $\mathbf{f}_{ctrl}(t,\bx,u)$  & $[45,52]\times[0,1]^4$   & $[0,1]$                \\ \midrule
		{$i=14$} & 9 & $\mathbf{f}_{rad}(t,\bx,u)$   & $[52,55]\times[0,1]^8$   & $[0,1]$                 \\ \midrule  
	\end{tabular}
\end{center}
\end{table}

\begin{table}
\begin{center}
	\begin{tabular}{@{}llll@{}} \toprule
		{Mode}  & $\tr = (\sti,\stj)$ & $\Guard_{\tr}$  &  $\Resetmap_{\tr}$        \\ \midrule  
		{$i=1$} & $(1,2)$  & $t == 4$ & $\begin{bmatrix} I_{5,5}\\ O_{4,4}\end{bmatrix}$          \\ \midrule  
		{$i=2$} &$(2,3)$   & $t == 7$ & $\begin{bmatrix} I_{5,5},O_{4,4}\end{bmatrix}$            \\ \midrule  
		{$i=3$} &$(3,4)$  & $t == 8$  & $\begin{bmatrix} I_{5,5}\\ O_{4,4}\end{bmatrix}$          \\ \midrule  
		{$i=\vdots$} & $\vdots$  & $\vdots$ & $\vdots$          								  \\ \midrule  
		{$i=13$} &$(13,14)$  & $t == 52$ & $\begin{bmatrix} I_{5,5}\\ O_{4,4}\end{bmatrix}$ \\ \midrule   
	\end{tabular}
	\caption{Transitions, guards, and reset maps of the controlled hybrid system $\Hybrid$.}\label{tab:trans}
\end{center}
\end{table}




For numerical reasons, it is necessary to scale the parameters and state variables, making it easier for the solver to succeed in solving the relaxed problem. Similarly, to facilitate the numerical optimization we rewrite the control variable $u(t) \in \InputSet = [0,1]$ as $u(t) = \zeta \hat{u}(t)$, with $\zeta\ll 1$ and  $\hat{u}(t) \in [0,1/\zeta]$. While the scale factor $\zeta$ may take different values depending on the numerical optimization details, the objective control $u(t)$ always evolves in $[0,1]$. 


%
Now that we have a valid controlled hybrid system, $\Hybrid$, we solve the optimal control problem with intermediate time points defined in \eqref{eq:Int_HOCP}, using the method from Section \ref{ssec:chs} and its implementation in Section \ref{ssec:implementation}.
The experimental measurement is modelled by the function $m(\bx) = x_6 + 4x_8$. Thus, we set
$
H(\bx(T_j)) := (x_6(T_j) + 4x_8(T_j) - z_{j})^2,
$
as we search to minimize the total residual error term:
\begin{equation}
	\varepsilon_{total} = \sum\limits_{1 \leq j \leq \nbexp}\frac{\sqrt{H(\bx(T_j))}}{\sum_{1 \leq j \leq \nbexp}z_{j}}\,.
\end{equation}

The original experimental data points $(T_j,z_j)$ are given in Table \ref{tab:datas}. 
\begin{table}[!ht] 
\begin{tabular}{lccccccc}
 \toprule
  Time ($h$)  & $7$ & $11$& $19$& $27$& $35$& $45$& $55$ \\
  Measure ($\frac{cpm}{1e^{-7}\textnormal{L}\cdot h^{-1}}$) & $16$ & $85$& $348$& $391$& $399$& $481$& $395$ \\
 \bottomrule
\end{tabular}
  \caption{Experimental data points $(T_j,z_j)$ used as references.} \label{tab:datas}
\end{table}
Here, the input control $u(t)$ models some hidden mechanism which evolves with the differentiation of the cells. 
It should be the same function of time for both the control and the radioactive cells batch. However, as the control generated by Algorithm \ref{alg:pw_ocp} is piecewise for each mode, and the fact that our data are on the radioactive species only, the solution of the optimization problem with only a final cost $H(\bx(T_i))$ is not \textit{balanced}, having a much stronger control in the modes where the radioactive species are evolving.  
\begin{figure}
	 \includegraphics[width = \columnwidth]{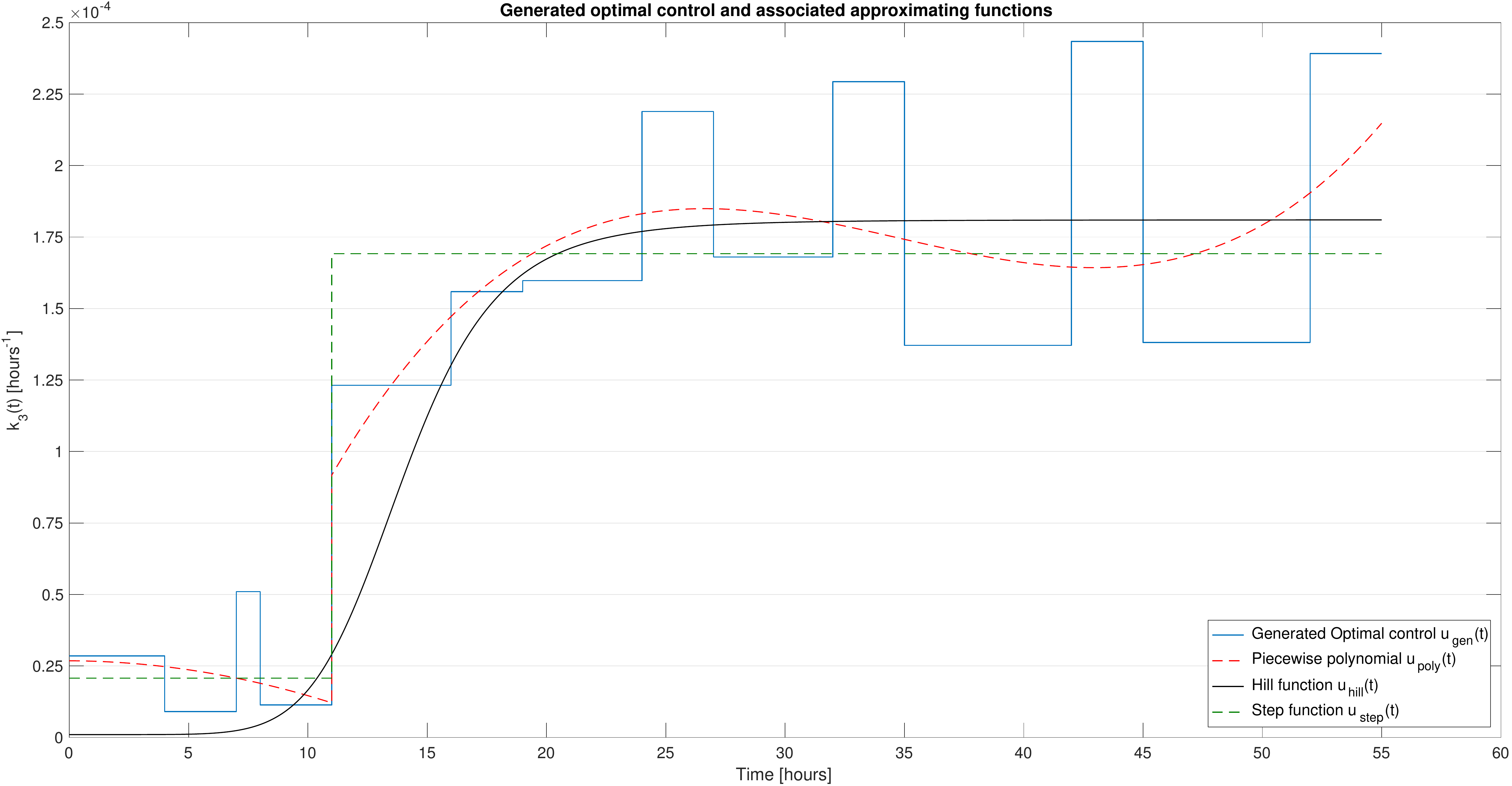}
	    \caption{Synthesised optimal control and various approximations that yield a realistic interpretation.}
    \label{fig:optimal_control}
\end{figure}

\begin{figure}
	\includegraphics[width = \columnwidth]{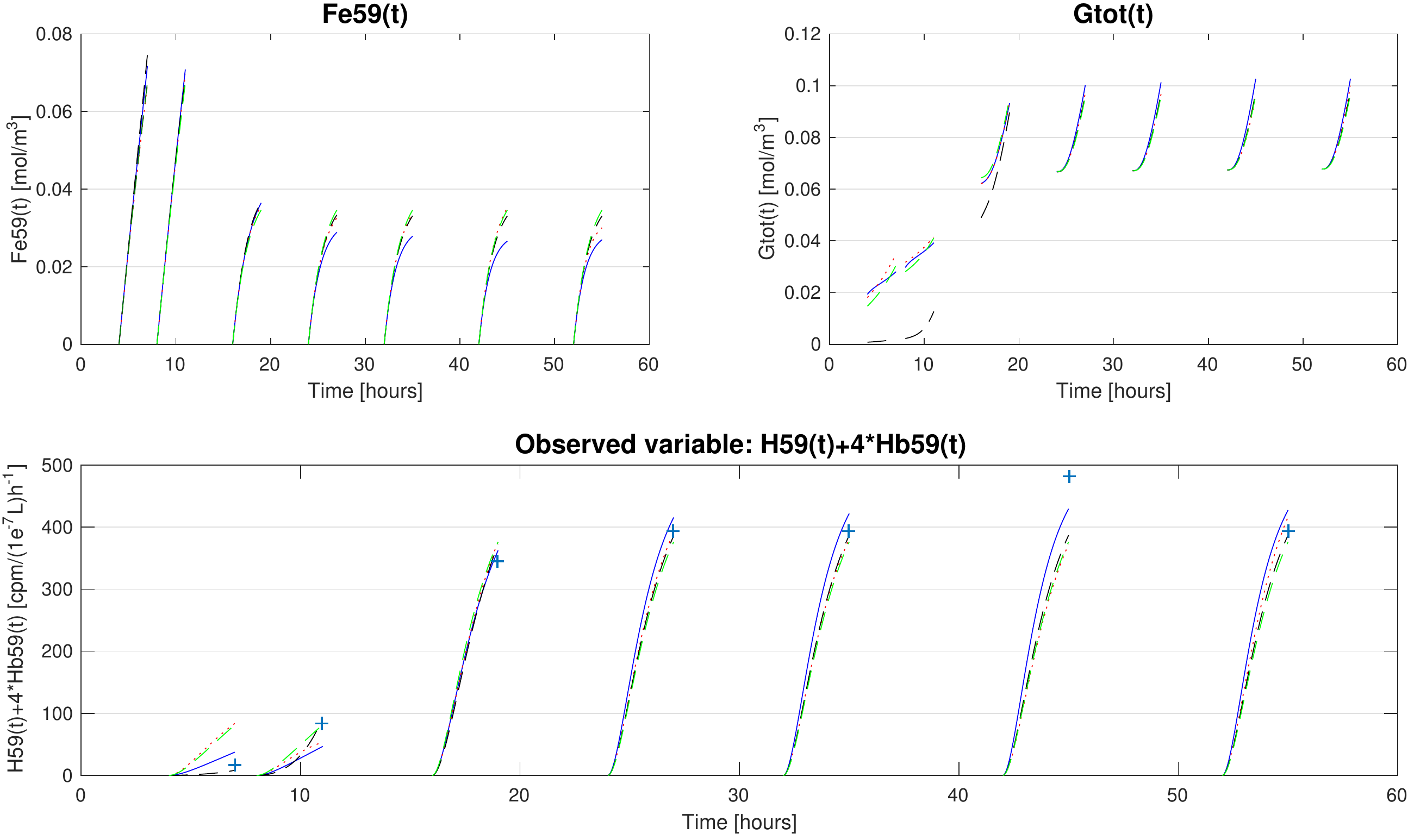}
    \caption{Radio-active variables $\ce{^{\scriptscriptstyle 59}Fe}$, $\ce{G}_{tot}$ corresponding to $x_5$ and $x_7$ in \eqref{eq:frad}, as well as, the comparison of the measurement function results to the data.}
    \label{fig:radioactive_variables}
\end{figure}
A workaround for the balancing problem is the following. 
We add a small penalization cost $c^1_i(t)  = (0.01u(t))^2$ to equilibrate the control when $i$ corresponds to a mode with radioactive species, otherwise $c^1_i(t) = 0$. 
In a similar vein, we add another penalization cost $c^2_i(t) = (u(T_j) - u(t))^2$ to avoid when the control strongly varies between two iterations $j$ on the interval $[T_{j-1},T_j]$ and $j+1$  on $[T_{j},T_{j+1}]$ (with the exception of the first iteration). 
This leads to $h_i(t,\bx(t),u(t)) = c^1_i(t) + c^2_i(t)$. 
Let us note that, even if these additional costs can eventually degrade the accuracy of the data fitting, we gain in terms of biological interpretation of the resulting traces.\\ 


Finally, by partitioning the computation in the time domain, we can greatly reduce the computational cost at each iteration. More technically, since the transitions of the hybrid system $\Hybrid$ are fully determined by the time $t$, we can pre-compute the function $\lambda: \R^+ \rightarrow \I$, which associates a mode $\lambda(t)$ to each time instant $t$. Thus, each iteration $j$ of Algorithm \ref{alg:pw_ocp} can be restrained to the sub-hybrid system $\Hybrid^j$ of $\Hybrid$, constituted by the modes visited in the interval $[T_{j-1},T_j]$.

For numerical implementation, the problem on measures is formulated in SPOTLESS\footnote{The SPOTLESS implementation was taken from https://github.com/spot-toolbox/spotless}, and then we extract the primal solution provided by a primal-dual SDP solver. To do so, we use the implementation from \cite{hybridOCP} to generate the dual problem. As an SDP solver we used MOSEK \cite{mosek} v.7.1. These tools are used in MATLAB v.9.0 (R2016a). Performance results are obtained with an Intel Core i7-5600U CPU ($2.60\, $GHz) with 16Gb of RAM running on Debian~8. 
We only solve the problem for a relaxation order $r=4$, as any higher order would be too memory expensive. 
In this particular example as constant input leaded to satisfying results, we did not impose any constraint $\epsilon$ in the algorithm.
Using this configuration, the total time taken by Algorithm \ref{alg:pw_ocp} is $2107$s, with $1700$s spent in the \texttt{HOCP} procedure, and $390$s in the \texttt{Synth} procedure. On Figure \ref{fig:optimal_control}, the control generated by Algorithm \ref{alg:pw_ocp} is shown in blue. 
This control is piecewise, and clearly divided in two phases: before and after $t$ equals $11$ hours. 
However, the control synthesised is still difficult to interpret as a biological phenomenon. Consequently, we propose three additional fits of this control to ease interpretation by using functions closer to biological knowledge. 
In Table \ref{tab:results} one can find the total error associated to all the possible controls, as well as the previous result of \cite{hemeModel}. 
In Figure \ref{fig:radioactive_variables}, we show a graphical representation of how closely each function can control the model to reach the desired data points.

\begin{table}[!ht] 
\begin{center}
\begin{tabular}{cc}
  \toprule
  Control Type  & $\varepsilon_{total}$  \\
  \midrule
  Original Control \cite{hemeModel} & $0.23$ \\
  Algo \ref{alg:pw_ocp} generated  & $0.096$\\
  Step function fit &  $0.12$ \\
  Piecewise Polynomial fit & $0.13$ \\
   Hill function fit &  $0.075$ \\
  \bottomrule
\end{tabular}
  \caption{Total error $\varepsilon_{total}$ associated to each proposed input.} \label{tab:results}
	\end{center}
\end{table}

\subsection{Discussion}

In a simulation-based approach, we have to propose for the desired time varying parameter, a template function to fit the data, e.g. a polynomial of given degree. If we want to fit a polynomial of higher degree, the simulations have to be run again multiple times. On the contrary, the proposed approach returns a control signal, and since the fit to data points is performed a posteriori, there is no additional computation cost in refining the model.

From the form of the experimental data points, an usual hypothesis is that $u(t)$ should be similar to a jump function, with a low value for the two first points, and a higher one for the following ones. However, even with such information a good fit is not easily achieved with simulations.
 
%

The control generated with Algorithm \ref{alg:pw_ocp} returns the expected ``jump'' behaviour for $u(t)$, and even with a low relaxation degree, the total residual error for the generated control is $9.59\%$ which is much lower than the $22.8\%$ from \cite{hemeModel}. 

We first fit a step function to the generated control, with a change at $t=11$. The associated error of $12.24\%$ is still lower than \cite{hemeModel}, yet being higher than the generated control mainly due to the second-to-last point.

The second fit is a piecewise polynomial function in two pieces. The first piece, for $t \in [0,11]$, is a polynomial of degree $2$ while the degree of the second, for $t\in[11,55]$, is $4$. This proposed input control allows to reproduce more accurately, than the step function, the third data point. However, its accuracy is worse on the first and two last points. The total error associated to this control is $13\%$, being overall the worst of the proposed fits.

Lastly, we fit a Hill function, a function used to model the kinetics of a class of biochemical reactions and which is a very common way to represent biological activation processes. The associated total error is $7.5\%$, which is the lowest, taking advantages from both the step function and the piecewise polynomial function. 
In this case, the inaccuracy also mainly comes from the second-to-last point, which is quite separated from trend of the other experimental points, and may be due to some experimental problems (no standard deviation results were available). Without taking this point into consideration for the error computation the error falls to $3\%$ for the Hill function fit.

On this particular example, this method provided a way to generate a control satisfying intermediate points without any \emph{a priori} on a particular form, avoiding the need for extensive numerical simulations. The generated control is accurate, and computed in a reasonable time ($\sim$35min), even for a large hybrid system of 14 modes with at most $9$ continuous variables. Using some fitting functions afterwards, it is even possible to refine the results and obtain a more interpretable function for the desired time varying parameters.

As, in this model, the sequence of transitions is known in advance, the use of \cite{hybridOCP} to solve \eqref{eq:hocp} at each iteration of Algorithm \ref{alg:pw_ocp} is arguable, as other methods can handle this problem. If needed  Algorithm \ref{alg:pw_ocp} can easily be adapted with another method to solve the optimal control problem on hybrid system, using for example simulations. However, Algorithm \ref{alg:pw_ocp} in its current form does not require any knowledge on the sequence of transition and can be applied to a larger set of biological models. 
 

 \section{Conclusion}
 \label{S:conclu}
 In this work, we address an important problem arising in biological modelling: model revision. We propose a method for revising an experiment modelled by a hybrid system, given a set of experimental data points. The method scales even on large hybrid systems such as the haemoglobin production model, while providing an accurate result, and a meaningful interpretation, as an activation process, for the mechanism underlying the revised parameter. The CHS formalism is motivated by the development of an automatic, and formal modelling of multiple-step experimental protocols, and to develop new methods for their analysis. 
Such formal representations had already been used as alternative, non-ambiguous languages, in contrast with the natural language, for the description of experiments \cite{exactLanguage}. However, those works do not consider an underlying mechanistic model in the form of ODEs. In a future work, we plan to provide an automatic and rigorous way to obtain a hybrid system modelling an experimental protocol with their underlying mechanisms. Moreover, we plane to investigate, using the CHS formalism, two other relevant problems in biological systems modelling: finding valid subsets of the parameters space fitting multiple data points (as an extension \cite{hybridBRS}), and the validation of biological experiments. The focus of this work has been on ODEs, but multicellular systems and transport processes are described by partial differential equation (PDE) models. The extension of semidefinite programming techniques to PDEs \cite{mevissen2011moment} and their application to biological models would require further theoretical and numerical developments. 





\bibliography{biosdp}

\section*{Appendix}
The parameters of the haemoglobin production model are shown in Table \ref{tab:params}. These are taken from the column $p_{mean}$ of \cite[Table 1]{hemeModel}. The ODEs used in the hybrid system $\Hybrid$ are \eqref{eq:fctrl} and \eqref{eq:frad}. The parameters $k_2$ and $k_7$ are set to $0$. The parameters $a_{k3}$ and $b_{k3}$ are not used, as we search a new function to model $k_3(t)$. The initial condition of the controlled hybrid system, $\Hybrid$, is determined by the initial condition in \cite{hemeModel}: $(i_0,\bx(0)) = (1,(0,\ce{Fe}_{0},0,0,0))$.
\begin{table}[!ht]
\begin{center}
\begin{tabular}{ccc}
  \toprule
  Parameter  & Numerical Value & Units  \\
  \midrule
  $k_1$ &  $1.4\num{10e-3}$  & $\textnormal{s}^{-1}$\\
  $k_4$ &  $4.47\num{10e-4}$ & $\textnormal{s}^{-1}$\\
  $k_5$ &  $7.27\num{10e-6}$ & $\textnormal{fL.molecules}^{-1}.\textnormal{s}^{-1}$\\
  $k_6$ &  $4.47\num{10e-4}$ & $\textnormal{s}^{-1}$\\
  $k_8$ &  $1.14\num{10e-5}$ & $\textnormal{s}^{-1}$\\
  $\ce{Fe}_{0}$ & $3.21\num{10e+2}$ & $\textnormal{atoms/fL}$\\
  
  $\ce{Fe}_{ex}$& $4$ & $\textnormal{atoms/fL}$\\
  
  $\ce{^{59}Fe}_{ex}$& $3000$ & $\textnormal{atoms/fL}$\\
  \bottomrule
\end{tabular} 
  \caption{The parameters (not scaled) for the ODEs $\mathbf{f}_{ctrl}$ and $\mathbf{f}_{rad}$, taken from the parameters set $pmean$ in \cite{hemeModel}.}\label{tab:params}
	\end{center}
\end{table}
\begin{equation}
	\tag{$\mathbf{f}_{ctrl}$}
	\label{eq:fctrl}
	\left\{
	\begin{aligned}
		\dot{x_c} = &\quad 1\\
		\dot{x_1} = &\quad k_1\ce{Fe}_{ex} - u(t)x_1(t)\\
		\dot{x_2} = &\quad - k_4x_2(t) -4k_5x_2(t)x_3(t) + u(t)x_1(t)  \\
		\dot{x_3} = &\quad k_6x_2(t) - 4k_5x_2(t)x_3(t)\\
		\dot{x_4} = &\quad k_5x_2(t)x_3(t) - k_8x_4(t)
	\end{aligned}
	\right.
\end{equation}

\begin{equation}
	\tag{$\mathbf{f}_{rad}$}
	\label{eq:frad}
	\left\{
	\begin{aligned}
		\dot{x_c} = &\quad 1\\
		\dot{x_1} = &\quad k_1\ce{Fe}_{ex} - u(t)x_1(t)\\
		\dot{x_2} = &\quad -k_4x_2(t) -4k_5x_2(t)x_3(t) + u(t)x_1(t)  \\
		\dot{x_3} = &\quad k_6x_2(t) - 4k_5x_2(t)x_3(t)\\
		\dot{x_4} = &\quad k_5x_2(t)x_3(t) - k_8x_4(t) \\
		\dot{x_5} = &\quad k_1\ce{^{59}Fe}_{ex} - u(t)x_5(t)\\
		\dot{x_6} = &\quad - k_4x_6(t) -4k_5x_6(t)x_7(t) + u(t)x_5(t)  \\
		\dot{x_7} = &\quad k_6(x_2+x_6)(t) - 4k_5(x_2(t)+x_6)x_7(t)\\
		\dot{x_8} = &\quad k_5x_6(t)x_7(t) - k_8x_8(t)
	\end{aligned}
	\right.
\end{equation}

\end{document}